\documentclass[referee]{aa} 
\usepackage{psfig}
\begin{document}

   \thesaurus{05     
              (09.03.1;  
               09.04.1;  
               13.07.1)  
               }
   \title{Dust and dark Gamma-Ray Bursts: mutual implications}


   \author{S.D. Vergani
          \inst{1},
          E. Molinari\inst{1}, F.M. Zerbi\inst{1}\and G.Chincarini\inst{2} \fnmsep
          }

   \offprints{S.D. Vergani}

   \institute{INAF-Osservatorio Astronomico di Brera, Via Bianchi 46 I-23807 Merate (Lc) Italy
\and
   Universit\`a di Milano Bicocca\\
              email: vergani@merate.mi.astro.it
    }

\authorrunning{Vergani, Molinari, Zerbi, Chincarini}
\titlerunning{Dust and dark GRBs}
   \maketitle

   \begin{abstract}

   In a cosmological context dust has been always poorly understood.
   That is true also for the statistic of Gamma-Ray Bursts (GRBs) so that
   we started a program to understand its role both in relation to
   GRBs and in function of z.

   This paper presents a composite model in this direction. The model
   considers a rather generic distribution of dust in a spiral galaxy and
   considers the effect of changing some of the parameters characterizing the
   dust grains, size in particular.
   We first simulated 500 GRBs distributed as the host galaxy mass distribution,
   using as model the Milky Way.
   If we consider dust with the same properties
   as that we observe in the Milky Way, we find that due to absorption we miss $\sim 10\%$
    of the afterglows assuming we observe the event within about
    1 hour or even within 100s.

   In our second set of simulations we placed GRBs randomly inside giants molecular
   clouds, considering different kinds of dust inside and outside the host cloud
   and the effect of dust sublimation caused by the GRB inside the clouds.
   In this case absorption is mainly due to the host cloud and the physical
   properties of dust play a strong role.
   Computations from this model agree with the hypothesis of host galaxies with
extinction curve similar to that of the Small Magellanic Cloud, whereas the host
cloud could be also characterized by dust with larger grains. Unfortunately,
 the present statistics lack solid grounds, being based on hardly compatible observations,
 at different time from the burst and with different limiting magnitudes.
 To confirm our findings we need a set of homogeneous infrared observations.
 The use of coming dedicated infrared telescopes, like REM,
   will provide a wealth of cases of new afterglow observations.

   \end{abstract}

%

\begin{table*}

\caption[]{Physical characteristics of H$_2$ galactic clouds.}
   \label{TabCloud}
  \[
    \begin{array}{p{0.15\linewidth}rlclclclclc}
     \hline
     \noalign{\smallskip}
     (1) & (2) &\hspace{0.5cm}& (3) &\hspace{0.5cm}& (4) &\hspace{0.5cm}& (5) &\hspace{0.5cm}& (6) &\hspace{0.5cm}& (7)\\
     \noalign{\smallskip}
     \hline
     \noalign{\smallskip}
      & R && n_H && \mathrm{No.(bulge)} &&\mathrm{No.(disk)} &&
\mathrm{Unit~mass} && \mathrm{Total~mass}  \\
     \noalign{\smallskip}
     \hline
     \noalign{\smallskip}
     GMC & 20 && 5\cdot10^2 &&   800 &&  4000 && 4\cdot10^5 && 1.92\cdot10^9\\
     DC  &  2 && 5\cdot10^4 && 16000 && 80000 && 5\cdot10^3 &&  4.8\cdot10^8\\
            \noalign{\smallskip}
            \hline
         \end{array}
      \]
\begin{list}{}{}
\item[] Radius is in pc and $n_H$ in cm$^{-3}$. Masses are in M$_\odot$.
\end{list}
\end{table*}

\section{Introduction}
The observations show that about 50\% of the detected GRBs are not visible at optical wavelengths.
 Statistics refers to well localized bursts
 (Bloom, Kulkarni \& Djorgovski 2002) with an X ray flux rather similar to those that have been
 detected also in the optical band. The rapid decline of the optical afterglow
 (Fynbo et al. 2002, Berger et al. 2002) can not explain the observations as well and
 that is why we call them dark bursts and an explanation has yet to be found.

In two cases, GRB 970828 and GRB 990506, in spite of lacking the optical afterglow we were
able to identify the host galaxies (z=0.958 and z= 1.3 respectively) so that in these cases
it is evident that high redshift is not the cause of the optical flux extinction.
A possible explanation is that the events have been obscured by dust.
 After the first part of our work was completed (Vergani 2002) Reichart (2001)
 and Reichart \& Price (2002) made a case for a strong absorption occurring in the molecular clouds
 where the event originates. This is in agreement with the current thinking that GRBs are
 related to massive star formation which is strongly correlated with very dusty regions.
 Whether we are dealing with a particular type of galaxies is not known.
 Ramirez-Ruiz, Threntham \& Blain 2002 assert that we might be dealing with ULIRG
 (Ultra Luminous Infrared Galaxies) or alike objects and that the extinction is
 essentially
 due to the dust distribution present in these galaxies. We do not have observational
 evidence that this is the case however. No matter how the dust is at work,
 we must also account for the local process of dust sublimation and understand how much
 dust a burst is capable of sublimating and sweeping out.

We decided to tackle the problem first theoretically,
and this paper report part of our work in this direction,
and observationally by building the robotic NIR Telescope REM
(Zerbi et al. 2002, Chincarini et al. 2003). The statistics of GRBs will also
largely increase as soon as the Swift satellite will be launched (Gehrels et al. 2003).

In section 2 we discuss how we model the dust itself, Section 3 is devoted to the
construction of the simple host galaxy; a basic GRB distribution models and
results are presented in section 4. In section 5 and 6 we associate GRBs with giant
molecular clouds and we explore its implications.
Conclusions are drawn in section 7.
\section{Dust model}
The afterglow radiation reaches the observer after interacting with circunburst
material, host ISM, IGM and the ISM of our galaxy. In this work we
consider only the interaction with the dust in the ISM of the host galaxy.

Our dust model is based on Mathis, Rumpl \& Nordsiek (1977) model, improved by Mathis
(1986). We suppose that dust is made by spherical grains composed by 50\% of graphite
and 50\% of silicates with a grain size $a$ distributed as

\begin{equation}
n(a)=n_{0}a^{-3.5}
\end{equation}

with $a_{min}=0.005\mu$m, $a_{max}=0.25\mu$m and $n_{0}$, proportional to the neutral hydrogen density
 ($n_H$[cm$^{-3}$]), used by \cite{vene} 2001.

The variation in magnitudes produced on flux after interacting with dust grains
is

\begin{equation}
\Delta m\propto\int_{0}^{\infty} Q_{ext} \pi a^{2}n(a)da
\end{equation}
with $Q_{ext}=$extinction efficiency (sum of scattering and absorption
efficiency, see \cite{Hulst}).

\begin{figure}
\psfig{figure=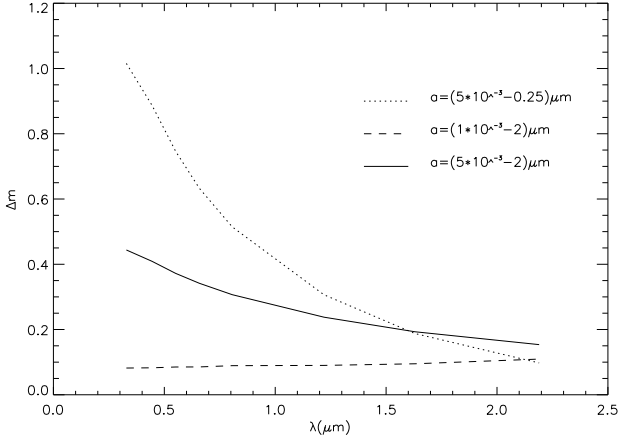,width=8.8cm}
\caption[]{Extinction curves produced by varying the dust grain size
in case of an hydrogen column density of $10^{21}cm^{-2}$.
Dotted steeper curve represents the extinction considering our Galactic-like
dust model (STD, see Tab.~\ref{TabRegions}) based on Mathis (1986).
Dashed and solid lines describe the extinction curve in case of a dust composed by larger grains
(OLG and LRG, respectively).}

         \label{figven1}
\end{figure}

By varying in our model the fraction of the two types of materials,
the exponent of the grain size distribution and the grain size, the result is a change in the scale of the
curve without modifying its shape. We cannot therefore disentangle their contribution from
different amounts of hydrogen column densities (NH[cm$^{-2}$]) along the line of sight.
On the contrary the size of the grains play a strong role. Large grains, which could be present in regions with intense star formation (\cite{Maio}) or in
circunburst environment after the burst has sublimated the smaller grains (\cite{vene}),
cause the extinction curve to flatten. For dust composed only by large grains ($a_{min}=1\mu$m, $a_{max}=2\mu$m) our
computations predict a fixed value of $\Delta m\sim0.1$, in agreement with the theory,
that predict for the case $2a>\lambda$ a fixed value of $Q_{ext}\sim2$
 (Fig.~\ref{figven1}). The effects of modifying $a_{min}$ with fixed $a_{max}$ are not
 relevant.

\section{Galactic-like model}

We model an host
galaxy similar to our Milky Way,
where the dust is distributed as the neutral hydrogen, both molecular
(H$_2$) and atomic (HI).

For the galactic hydrogen surface density we adopt the values
taken from \cite{encyscoville} and \cite{BM}.

\begin{figure}
\psfig{figure=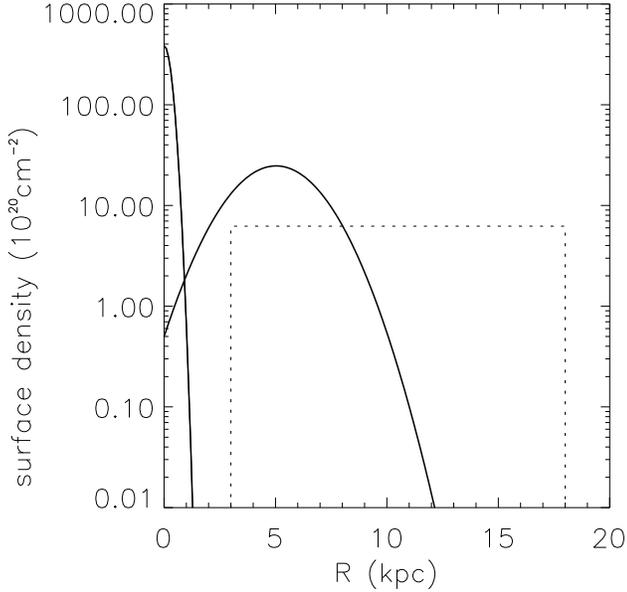,width=8.8cm} \caption[]{Milky Way hydrogen
column density as a function of radial distance R in kpc. The two
solid curves are the two gaussian fits to the distribution
reported in \cite{encyscoville} for the molecular hydrogen
(H$_2$). Dashed line is the atomic hydrogen (HI) surface density,
simplified as a box function between 3 and 18 kpc from galactic
center.}

         \label{FigGauss}
\end{figure}

For H$_2$, two gaussian fits were drawn on nuclear and disk data:
namely the gas density has been expressed as:
\begin{equation}
 \mathrm{ NH (bulge)}\propto e^{-{R^2}\over{2(0.28)^2}}
\mathrm{cm}^{-2}
\end{equation}
and
\begin{equation}
 \mathrm{ NH (disk)}\propto e^{-{(R-5.03)^2}\over{2(1.8)^2}}
\mathrm{cm}^{-2}
\end{equation}
for the bulge and the disk, where $R$ is the radial distance in cylindrical
coordinates (kpc).
The total mass of galactic H$_2$ is distributed with $4\cdot10^8M\odot$ in the
bulge and $2\cdot10^9M\odot$ in
the disk. The H$_2$ clouds are present in two distinct
morphologies, which we catalogue as giant
molecular clouds (GMC) and dense clouds (DC) whose adopted characteristics are
summarized in Table \ref{TabCloud}.
An algorithm assigns positions, with a random distribution
weighted by Galaxy mass density, to GMC and DC.

\begin{figure}
\psfig{figure=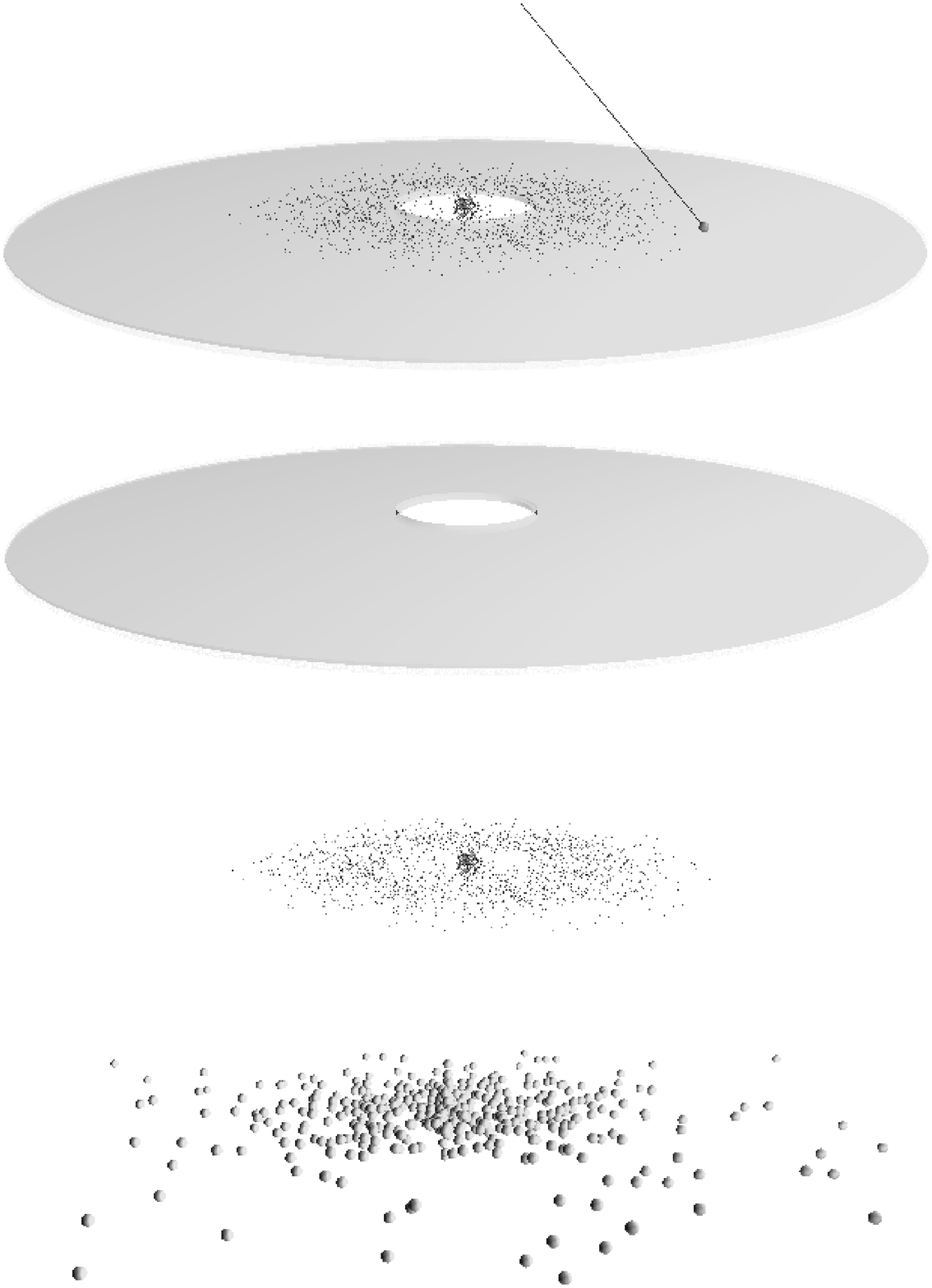,width=8.8cm}
\caption[]{A rendered image of our host galaxy model. The top figure is
the complete model with the sphere on the right which represents
a randomly positioned GRB. Its line of sight crosses the galactic
plane. Below the distribution of the atomic hydrogen (HI) and  H$_2$
clouds (both GMC and DC) are separately shown.
In the bottom picture the simulated distribution of the 500 GRBs.}
\label{FigPov}
\end{figure}

Neutral atomic hydrogen is present in a diffuse form throughout
the disk system with a total mass of $4.3\cdot10^{9}\odot$.
Following \cite{BM}, we simplify its distribution as
being sharply confined in the radial interval $3<R<18$ ($R$ in
kpc) with a spatial density represented by $n_H[\mathrm{cm}^{-3}]
= 0.797 e^{-h^2/0.02}$ (where $h$ is the height from Galactic plane in kpc). In Fig. \ref{FigGauss}
we plot the adopted fits for HI
(dashed line) and H$_2$ (solid lines) surface densities, while Fig. \ref{FigPov} shows the appearance of
the host galaxy hydrogen distribution .

\section{Basic GRBs absorption model}

\begin{figure*}
\psfig{figure=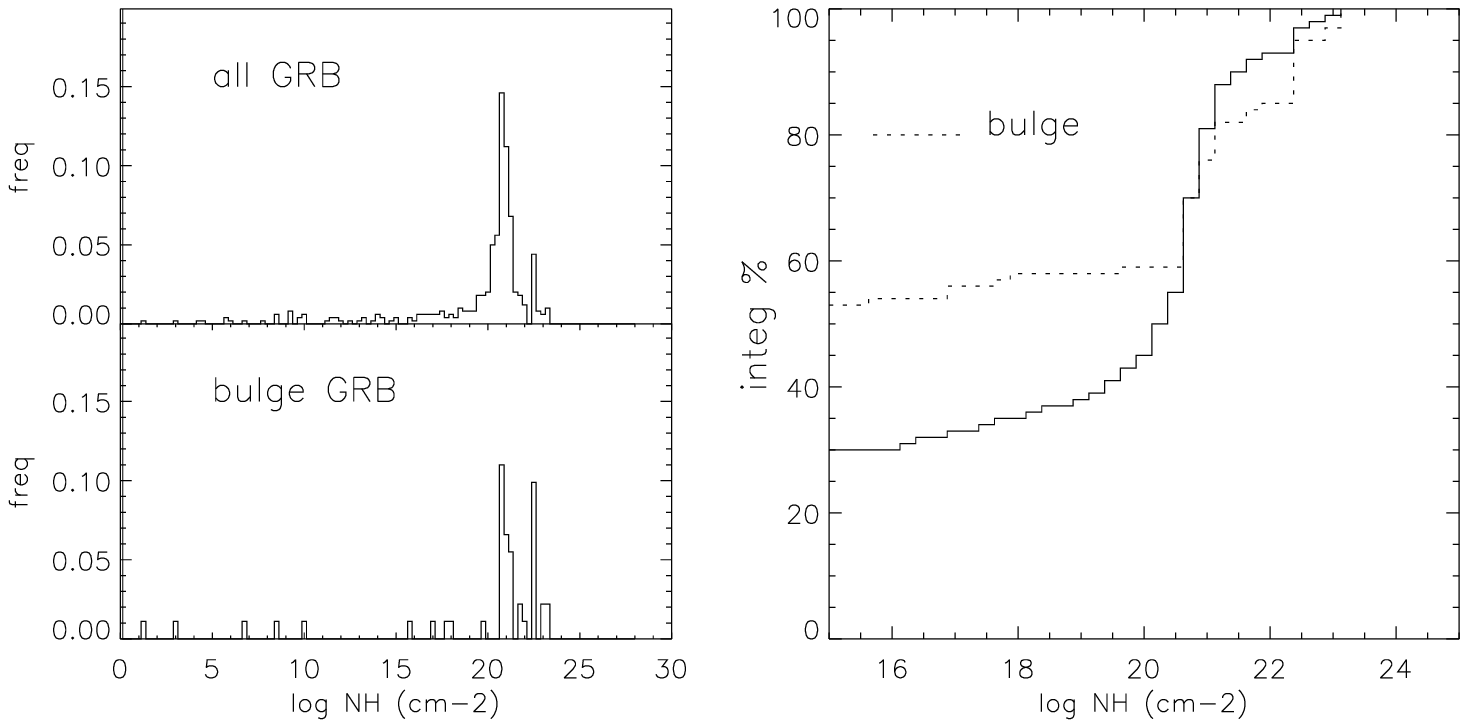,width=18cm}
\caption[]{Hydrogen column density distribution for 500 simulated GRBs.}

         \label{fignh}
\end{figure*}

We suppose that the GRBs are
distributed as the luminosity (barionic mass) of the host Milky Way-like galaxy. Using the Milky
Way photometric model of \cite{KDF}, we have

\begin{equation}
j_{d}(R,h)=\frac{I_{d}}{2h_{0}}e^{-R/R_{d}-|h|/h_{0}},
\end{equation}

with

\begin{equation}
h_{0}=\left[0.165+0.21(R/R_{0}-\frac{5}{8})\right]\textrm{kpc},
\end{equation}

$R_{d}=$3kpc, $R_{0}=$8kpc and the central surface brightness
$I_{d}=$1000
$L_{\odot}pc^{-2}$.

The bulge has been modeled using a King profile
\begin{equation}
I(r)\propto[1+(r/r_{c})^{2}]^{-3/2}
\end{equation}

with $r_{c}=$0.25kpc.

The fraction of GRBs located in the Galactic disk and in the bulge
is derived assuming a mass ratio disk to bulge of 4.5.

The burst occurs randomly using as weighting function the distribution of
mass of the galactic model. The observer is located randomly over a $4\pi$ solid angle.

Each molecular hydrogen cloud is accounted for absorption
and the amount of NH is integrated along the line of sight.
The quantized NH amount is then $4.5\cdot10^{22}$cm$^{-2}$ for the GMC and
$2.25\cdot10^{23}$cm$^{-2}$
for the DC.

The atomic diffuse hydrogen is summed integrating along the
line of sight from the GRB to a radius of 20 kpc from the galactic
center. The HI
contribution ranges therefore from a null value to $7\cdot10^{22}$cm$^{-2}$,
with a pick at about $5\cdot10^{20}$cm$^{-2}$ not sufficient to obscure the afterglow.
On the other hand the encounter of a single cloud yields values of NH
of about $5\cdot10^{22}$cm$^{-2}$ (GMC) or $5\cdot10^{23}$cm$^{-2}$ (DC)
largely enough to completely absorb the optical afterglow.

The total amount of NH is computed and catalogued for the whole 500
GRBs set and for the subset of
 91 nuclear bursts. In Fig.~\ref{fignh} the histogram of the total NH
distribution is shown for the
two populations. To better appreciate the difference a cumulative
distribution  for a smaller range of column densities is reported aside.

\begin{table*}
      \caption[]{Extinction limits computed on the basis of the limiting
      magnitudes of different instruments and of GRB990123 afterglow light curve.  NH values are computed using the dust model in Section 2.}
         \label{TabLim}
      \[
         \begin{array}{p{0.15\linewidth}lclclclclclclclcl}
            \hline
            \noalign{\smallskip}
            (1) & (2) &\hspace{0.45cm}& (3) &\hspace{0.45cm}& (4) &\hspace{0.45cm}& (5)
            &\hspace{0.45cm}& (6) &\hspace{0.45cm}& (7) &\hspace{0.45cm}& (8) &\hspace{0.45cm}& (9)
             &\hspace{0.45cm}& (10)\\
            \noalign{\smallskip}
            \hline
            \noalign{\smallskip}
             &  \mathrm{sec} &&  m_R && m_K &&  R_{\mathrm{lim}} &&
K_{\mathrm{lim}} &&  A_R && A_K &&  NH_R && NH_K\\

            prompt & 100 &&  10 && 7.5 &&  19 && 15.5 && 9 && 8
            && 0.86 && 2.13
            \\

           late  & 5000&&  16 && 13.5 &&  24 && 20.5 &&  8 && 7
           && 0.76 && 1.87
            \\

            \noalign{\smallskip}
            \hline
         \end{array}
      \]
\begin{list}{}{}
\item[(3)(4)] In these columns are reported light curve values GRB990123-like.
\item[(5)(6)] Prompt observation simulated with REM (\cite{tesusi}, Table 2.3),
late observation simulated with ISAAC and FORS (see
http://www.eso.org/oserving/etc/ for its ETC time estimates)
\item[(9)(10)] in units of $10^{22}\mathrm{cm}^{-2}$
\end{list}
\end{table*}

\begin{table*}
 \caption[]{Optical (R) and infrared (K) transient loss (\%). Prompt and late
      observation cases are reported for z=1 galaxy position.}
   \label{TabPersi}
  \[
    \begin{array}{p{0.04\linewidth}p{0.15\linewidth}clclclc}
     \hline
     \noalign{\smallskip}
     \multicolumn{2}{c}{(1)} & (2) &\hspace{0.6cm}& (3) &\hspace{0.8cm}& (4) &\hspace{0.8cm}& (5)
      \\
     \noalign{\smallskip}
     \hline
     \noalign{\smallskip}

     &&\multicolumn{3}{c}{\hrulefill~\mathrm{\% ~lost ~of ~all ~GRBs}~\hrulefill}
     && \multicolumn{3}{c}{\hrulefill~\mathrm{\% ~lost ~of ~bulge ~GRBs}~\hrulefill} \\
     &&  R && K &&  R && K  \\
     \noalign{\smallskip}


     \hline
     \noalign{\smallskip}
     z=1 & prompt&   9 && 7 && 18 && 14 \\

         &  late&   9 && 7 &&  18 && 14 \\

            \noalign{\smallskip}
            \hline
         \end{array}
      \]

\end{table*}

\begin{figure}
\psfig{figure=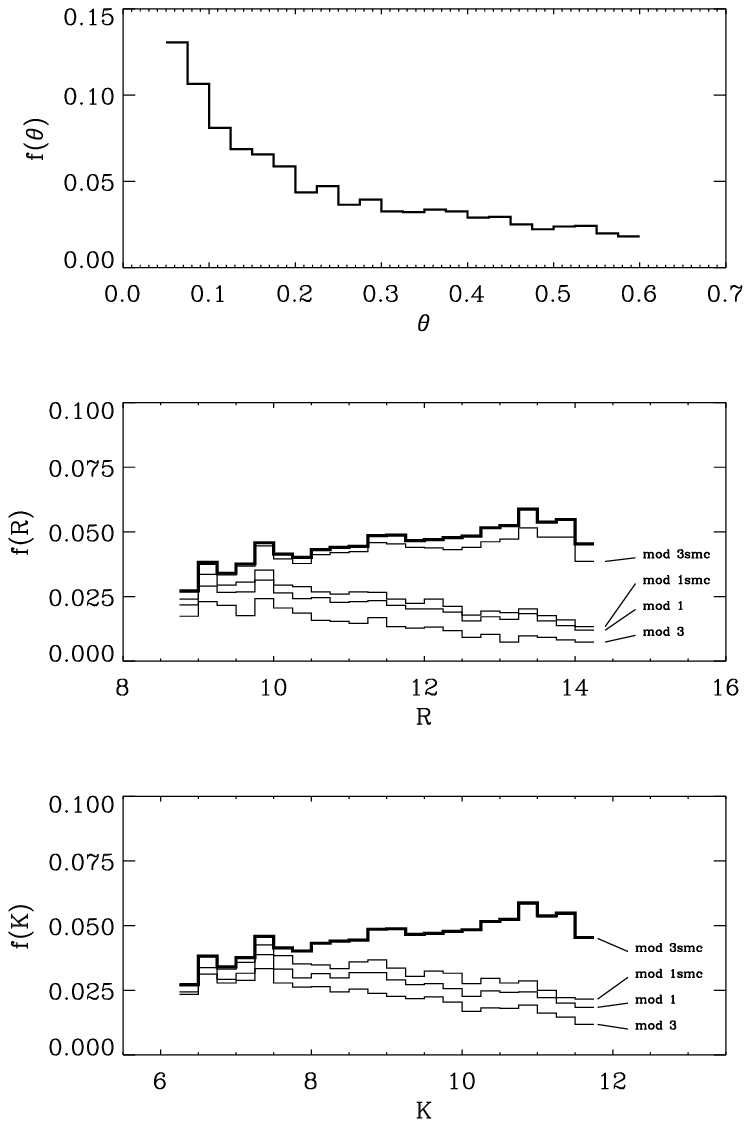,width=8.8cm} \caption[]{ Distribution of the
$\theta$ jet opening angles of our 5000 GRBs
simulated sample (upper panel). The parent distribution follows a
$\theta^{-0.85}$ law after a fit to
observed jet angles. The middle panel shows the translation of
jet angles in R magnitudes (heavy
line). The transformation lays on the zero point of a GRB990123-like
event at z=1 ($\theta = 0.086$,
R=10, K=7.5) and the assumption of the conservation of the total flux
on the emitting solid angle
($\propto \theta^2$). Thinner lines are the luminosity function of the
promptly observed (non absorbed in
our simulations) afterglows in the R band. The four histograms refer to
different models: 1, 3, 1smc and
3smc. In the lower panel the same set of distributions are plotted for
the K filter (note that model 3smc
does not differ from the original distribution, plotted as a heavy
line). } \label{figteta}
\end{figure}

The amount of dark GRBs due to dust absorption can now be
estimated, once we have
fixed the limiting magnitude of our telescopes in the various passbands
and the typical
apparent magnitude of GRB afterglows.
We associate randomly to each GRB a jet opening angle $\theta$
following the law $\propto\theta^{-0.85}$ that we have extrapolated
from the known jet angles reported by \cite{Bloomb} (see Fig.\ref{figteta}, upper panel).
Note that the $\theta$ distribution is truncated as $0.05<~\theta[{\mathrm rad}]<0.6$.
Considering the luminosity L proportional to
$\theta^{-2}$, we calculate the magnitudes R and K of GRBs at 100s and 5000s, taking as
reference GRB990123 shifted at z=1 (that is the observed mean redshift of GRBs,
\cite{Hurley}), the color data by \cite {Simon} and using the relation

\begin{equation}
 m_{GRB}=m_{990123}+5\log(\theta/\theta_{990123})
\end{equation}

where $\theta_{990123}$ and $m_{990123}$ are the jet angle and the magnitudes
of GRB990123.
To compute the limiting magnitudes we assume to use REM telescope for the fast
response to the GRB alert in R and K bands,
FORS and ISAAC camera of ESO/VLT for the long term observations (Tab.~\ref{TabLim}).

For each GRB we calculate the extinction in R and K caused by the traversed NH column
density. We are then able to compute the percentage of lost afterglows
due to dust absorption
under the assumption that the host galaxy at z=1 has
dust properties similar to Milky Way.

In Table~\ref{TabLim} R and K magnitudes of prompt and late observations for
GRB990123-like events are reported. The values of NH quoted in columns (9) and
(10) represent the computed amount of NH needed, according to our dust model,
to obtain the extinction values of column (7) and (8).
The summary of Table~\ref{TabPersi} is a clear indication that only a
relatively small fraction of GRBs
is {\it not} observed due to dust extinction in the case of a host galaxy
at z=1.

\section{GRBs in molecular clouds}

\begin{figure}
\psfig{figure=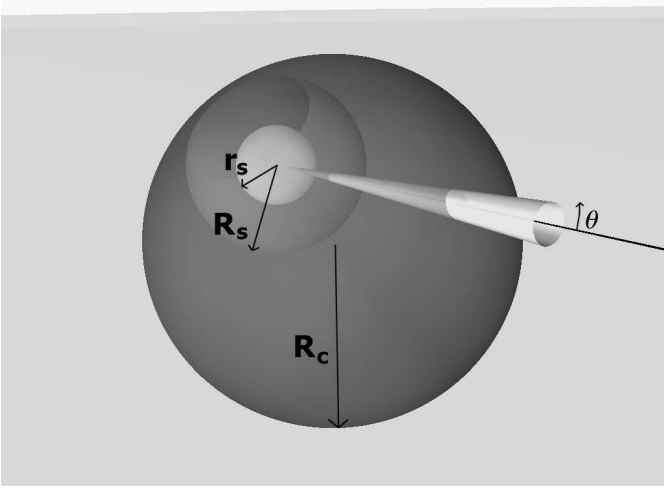,width=8.8cm} \caption[]{ Geometry around
the GRB location explaining the
shapes of the four regions whose content in terms of dust is
parameterized in our models. The regions are
spherical shells centered on the GRB location delimited by  $r_s$(the
radius up to which all dust grains are destroyed),
$R_s$(the
radius up to which grains with radius smaller 1$\mu$m are destroyed),
$R_C$ (the radius of the giant molecular cloud) and the border of the galaxy.
$\theta$ is the jet opening
angle associated to the GRB event.}
\label{figregion}
\end{figure}

Our results show that, if the GRBs distribution is similar to the mass distribution
 and if the dust of the host galaxies has the same characteristic of
Galactic dust, the percentage of dark GRBs due to dust obscuring
is rather low. There is no relevant difference between R and K observations
and between percentages relatives to observations taken at 100s and 5000s.
It also happens that no burst, out of the 500 considered,
occurs statistically inside a molecular cloud.
 In this scenario the majority of dark bursts could be due to
high redshift Ly-$\alpha$ absorption.

 We then consider the case that GRBs follow the distribution of giant molecular clouds.
 Physically the assumption is that of
 a strong connection between GRBs and massive stars
formation.

Within this framework, we place 5000 GRBs
inside our modeled giant
molecular clouds, that have similar characteristics respect the supposed typical GMC GRBs
host described by \cite{Galama}.

 The first aspect to consider is dust sublimation by the optical-UV flash accompanying the GRB,
phenomenon likely confirmed by the observations of the afterglow light curve of GRB
990123 (\cite{Akerlof}). We consider the results of \cite{Wax} and
\cite{Reichsub} that compute the radius up to which the dust is sublimated,
which is also a function of dust grain size.
In case of a canonical distribution of graphite and silicate grain size, the
sublimation radius is
$R_s\simeq 10 L_{49}^{1/2}pc$ where $L_{49}$ is the 1-7.5eV (1600-12000\AA) isotropic-equivalent peak luminosity of the optical
flash in unit of
$10^{49}erg s^{-1}$, that is the 1-7.5eV isotropic-equivalent peak luminosity of the optical
flash of GRB990123.

Inside the host cloud we consider both the case of a standard galactic dust and the case of
a dust, already present before the sublimation took place, with larger grain size with
 $a_{min}=0.005\mu$m, $a_{max}=2\mu$m, because the connection of GRBs
and intense star formation regions (\cite{Galama}), as described in $\S2$.
Moreover, in the case of large grains, we have to consider two different
sublimation radius: an inner one ($r_s$) up to which all grain are destroyed
and a larger one ($R_s$) up to which only the grains whit radius smaller than
$1\mu$m can be sublimated (\cite {vene}).

$R_s$ and $r_s$ vary with the intensity of the peak luminosity of the optical
flash, that we can suppose depending on the GRB jet opening angle $\theta$. We calculate for each GRB $R_s$ and $r_s$ as $R_s=R_{s990123}*\theta_{990123}/\theta$
and $r_s=r_{s990123}*\theta_{990123}/\theta$, where $R_{s990123}$,
$r_{s990123}$ and $\theta_{990123}$ are the two sublimation radius and the jet opening
angle of GRB990123.

To each of the 5000 simulated GRBs is associated a random line of sight passing
through 4 regions: $\Re1$ is the inner
region from the GRB to $r_s$ (the
radius up to which all dust grains are destroyed); $\Re2$ goes from
$r_s$ to $R_s$, where large grains
(radii from 1$\mu$m to 2 $\mu$m) are allowed to survive;
$\Re3$ is the rest on undisturbed
host cloud dust;
$\Re4$ is the host galaxy outside
the host cloud, consisting of all other molecular clouds and of the
diffuse medium (see Fig.\ref{figregion}). In these regions we place 5 kind
of dust yielding ten different models.

In the first three models, to calculate the extinction in the region outside
the host molecular clouds, we use our Galactic-like dust model.
In $\Re3$, that is the part inside the host molecular clouds where there is no
sublimation, 2 scenarios are explored: standard galactic dust (model 2 and 3) or a similar
one but with larger grains ($a_{max}=2\mu$m, model 1). Moreover, in model
2 we suppose that dust in the whole cloud is standard and thus sublimated both
in $\Re1$ and $\Re2$, whereas in model 1 and 3 the circunburst medium is supposed to
be characterized by dust with $a_{max}=2\mu$m, thus in $\Re2$ dust grains larger than
$1\mu$m
are left.

Hyorth et al., 2003 have recently supported the hypothesis that GRBs environment has an extinction
curve similar to the one of Small Magellanic Cloud (SMC),
so, in model 1smc, 2smc and 3smc
we present the same scenario of model 1, 2 and n 3 but considering SMC extinction
 inside the cloud and/or in the host galaxy.
Both the cases of extinction caused by larger grains or SMC-like dust agree
 with the non detection of the $\lambda=2175\AA$ bump (\cite{Galama}), that is
 instead a significant feature of Galactic-like dust.

Simulations by Perna et al. 2003, show that if we consider dust with
properties similar to the Galactic one, burst energy sublimes very quickly
silicate grains present in the circunburst medium differently from graphite
made grain that need a longer time to be destroyed and whose sublimation is
more efficient on the smaller grains. On this basis, we think reasonable
to consider models (model 1b, 3b, 1smcb and 3smcb)
in which dust in $\Re2$ region is composed only by graphite
grain with $a_{min}=1\mu$m, $a_{max}=2\mu$m.
 Table \ref{TabRegions} shows all
the cases.

\begin{table}
 \caption[]{Regions around GRB location and their contents in term of
dust type as adopted in
 our computations. Dust model used are: STD (our model of standard Galactic dust,
with grain radii from 0.005$\mu$m to 0.25$\mu$m);
 LRG (with added larger grains, from 0.005$\mu$m to 2$\mu$m); OLG (only large
grains, left from partial sublimation,
 with radii from 1$\mu$m to 2$\mu$m); GRA (only large graphite
grains, left from partial sublimation)  SMC (dust with low extinction similar to
Small Magellanic Clouds); SUB (no dust,
 completely sublimated by the burst).
  }
   \label{TabRegions}
  \[
    \begin{array}{p{0.31\linewidth}clclclc}
     \hline
     \noalign{\smallskip}
     (1) & (2) &\hspace{0.2cm}& (3) &\hspace{0.2cm}& (4)
&\hspace{0.2cm}& (5)\\
     \noalign{\smallskip}
     \hline
     \noalign{\smallskip}
     &  \Re1 && \Re2 && \Re3 && \Re4 \\
     \noalign{\smallskip}
     \hline
     \noalign{\smallskip}
      model 1       & SUB && OLG && LRG && STD \\
      model 1b      & SUB && GRA && LRG && STD \\
      model 2       & SUB && SUB && STD && STD \\
      model 3       & SUB && OLG && STD && STD \\
      model 3b      & SUB && GRA && LRG && STD \\
      model 1smc    & SUB && OLG && LRG && SMC \\
      model 1smcb   & SUB && GRA && LRG && SMC \\
      model 2smc    & SUB && SUB && SMC && SMC \\
      model 3smc    & SUB && OLG && SMC && SMC \\
      model 3smcb   & SUB && GRA && SMC && SMC \\
     \noalign{\smallskip}
     \hline
     \end{array}
   \]
\end{table}

\begin{table}
 \caption[]{Optical (R) and infrared (K) transient loss (\%) in the GRB-GMC association scenario. The
 models are described in text and Tab. \ref{TabRegions}.
 }
   \label{TabPersi2}
  \[
    \begin{array}{p{0.22\linewidth}clclclc}
     \hline
     \noalign{\smallskip}
     (1) & (2) &\hspace{0.6cm}& (3) &\hspace{0.8cm}& (4) &\hspace{0.8cm}& (5)\\

     \noalign{\smallskip}
     \hline
     \noalign{\smallskip}
     &\multicolumn{3}{c}{\hrulefill~R~\hrulefill}
     &&\multicolumn{3}{c}{\hrulefill~K~\hrulefill}\\
     &  \mathrm{prompt} && \mathrm{late} &&  \mathrm{prompt} && \mathrm{late} \\
     \noalign{\smallskip}
     \hline
     \noalign{\smallskip}
     model 1       & 53.0 && 57.3 && 38.8 && 44.8 \\
     model 1b      & 52.8 && 57.1 && 38.5 && 44.5 \\
     model 2       & 67.9 && 69.8 && 49.5 && 54.0 \\
     model 3       & 68.2 && 70.0 && 49.9 && 54.5 \\
     model 3b      & 57.8 && 57.1 && 38.5 && 44.5 \\
     model 1smc    & 46.8 && 51.1 && 29.9 && 36.6 \\
     model 1smcb   & 46.6 && 50.8 && 29.4 && 36.4 \\
     model 2smc    &  7.0 && 10.4 &&  0.2 &&  0.3 \\
     model 3smc    &  7.2 && 10.7 &&  0.2 &&  0.4 \\
     model 3smcb   &  7.1 && 10.6 &&  0.2 &&  0.3 \\

     \noalign{\smallskip}
     \hline
    \end{array}
  \]
\end{table}

\begin{figure}
\psfig{figure=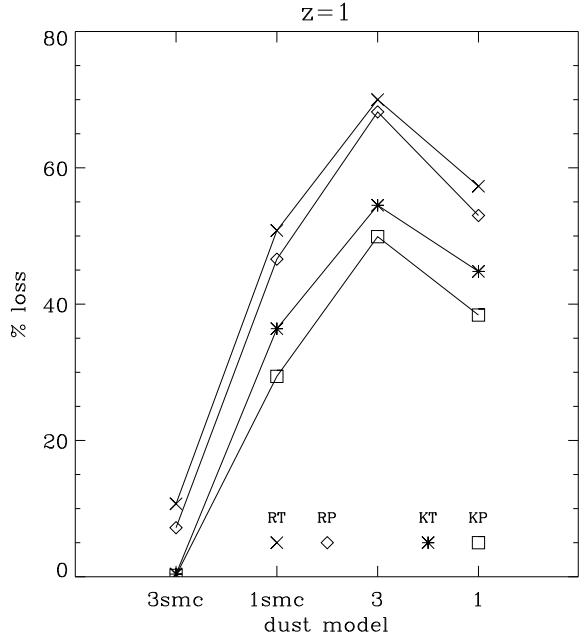,width=8.8cm} \caption[]{Trends for the most
relevant models of the percentage for prompt (RP, KP)
and late (RT, KT) observations in R and K bands of
dark bursts caused by dust extinction.}
\label{durezza}
\end{figure}

\section{Results}

The percentage of dust obscured afterglows for each model is computed summing all the absorption
intervening in the four regions. Critical limiting magnitudes, beyond which the transient is no more visible,
are taken from columns (5) and (6) of Tab.~\ref{TabLim} while extinction curves used for standard Galactic
dust and larger grain dust are respectively the dotted and the solid ones represented in
Fig.~\ref{figven1}. For SMC extinction curve we use the one reported by \cite{SMC}.

 If in $\Re2$ there are only grains larger than 2$\mu$m
 (both in case of standard and graphite only grains) we use
 dashed curve of Fig.~\ref{figven1}, but multiplied respectively by a factor
 $F_{1}=\int_{1}^{2}a^{-0.5}da/\int_{0.005}^{2}a^{-0.5}da$ and $F_{2}= 0.5*F_{1}$ taking into account that total
 dust mass is decreased. The 0.5 factor takes into account that half of the
 dust population (silicates) has completely disappeared.

 In Tab.~\ref{TabPersi2} results for different models are reported.
 It is evident that content in $\Re2$ is almost not significant, whereas
most of the extinction take place in $\Re3$, so the content of this region has
a determinant role.
In Fig.~\ref{durezza} we plot percentages for relevant cases.

In the lower
 panels of Fig.~\ref{figteta}, we show magnitudes distributions (R and K) of
 the simulated afterglows after 100s for GRBs at z=1. Together with the
 complete population, the distributions of observed transient for the different models are
 plotted.

\begin{figure}
\psfig{figure=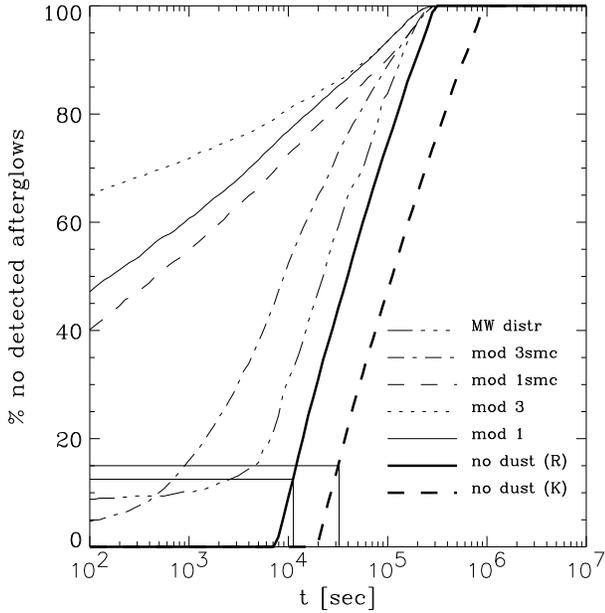,width=8.8cm} \caption[]{Percentage of lost afterglows
in function of observing time for GRB at z=1 with our simulated light curves.
Solid bold line represents the R band curve without considering dust extinction,
$R_{lim}=21$; dashed bold line represents the K band curve without considering dust extinction, $K_{lim}=19$,
whereas the other curves report R band results from our models calculated for $R_{lim}=21$.
 Two line intersections show the percentages in R and K relatives to the data of GRB020819.}
\label{senzaz1}
\end{figure}

A comparison of our simulation thought in view of Swift and REM
with observational data available now would be misleading.
 From our point of view it is practically impossible to extract a
statistically significant number (e.g. \% of real dark bursts) from a set of observations performed with
different telescopes (i.e. limiting magnitudes) and at different time from the
burst. To stress this point we compute at different times the percentage of events for which
the OT would not be observed even without dust extinction, considering the
light curves of our simulated GRBs and fixing R limiting magnitude of
$R_{lim}=20.5$. Solid bold line in Fig.~\ref{senzaz1} shows that for the majority
of data observational times the percentage of lost events is significant.
Dashed bold line of Fig.~\ref{senzaz1} reports the computation made for the K
infrared transient with $K'_{lim}=19$.

Recently \cite{Klose} have dealt with a GRB (GRB020819) whose afterglow has
been searched without success in K', 0.37 days after the burst down to K'=19
 and in R band 0.13 days after the burst down to R=20.5. We report these observational times and magnitudes
  in Fig.~\ref{senzaz1}: we notice that we can not exclude that
 GRB020819 could not be a real obscured or high redshift burst.
 It is also worth noting that our computations remark that if the observations reach $R_{lim}=24$, the
 value at 0.13 days of the R band curve not affected by dust is null
 and that late K observations
 (more than 6 hours after the burst) are always sensibly biased.

 We add to Fig.~\ref{senzaz1} the R band results of our models
 (i.e. considering extinction) at the same
 magnitude limit.

\section{Conclusions}

We have considered different kinds of dust and their role in obscuring optical
and NIR afterglows of GRBs simulated with different distributions inside a
galaxy at z=1.
In the case of GRBs occurring inside giant molecular clouds,
 our simulations show that dust in GRBs host galaxies has not the same properties of
galactic dust, otherwise dark GRBs would be more than observational
data say. Furthermore our results agree with the hypothesis of dust extinction curve of
GRBs host galaxies similar to that of SMC. Moreover, if the host
molecular cloud is characterized by large dust grains, high redshift plays a
minimal role in the causes of dark GRBs. The opposite
 situation takes place if
also inside the host molecular cloud dust has SMC properties.

Once we will gather enough prompt K observations we will be able to estimate
the nature of dust present in host molecular clouds from the fraction of lost K
afterglows.

We underline the fact that with present observational data it is impossible to
produce a real statistic for dark bursts due to the late and varying time of
observations and to the different magnitudes reached.

In the future, from the afterglow results given by Swift and
REM, we will be able to determine the nature of dark GRBs.
Indeed with a further refinement of the model and a good statistics
of bursts observed in various colors we will certainly be able to know
which kind of dust is present in the environment of the burst.

\begin{acknowledgements}
      The authors want to thank Stefano Covino and Daniele Malesani for the
      useful discussions.
\end{acknowledgements}

\end{document}